\documentclass[twocolumn,prb,floatfix,amsmath]{revtex4}
\usepackage{graphicx}

\begin{document}

\title{ Lithiation of InSb and Cu$_2$Sb : A Theoretical Investigation }

\author{S. Sharma}
\email{sangeeta.sharma@uni-graz.at}
\author{J. K. Dewhurst}
\author{C. Ambrosch-Draxl}
\affiliation{Institute for Theoretical Physics, Karl--Franzens--Universit\"at Graz,
Universit\"atsplatz 5, A--8010 Graz, Austria.}

\date{\today}

\begin{abstract}
In this work, the mechanism of Li insertion/intercalation in the anode materials InSb and Cu$_2$Sb is 
investigated by means of the  first-principles total-energy calculations. The electron localization 
function for the lithiated products of InSb are presented. Based on these results the change in the 
bonding character on lithiation is discussed. Further, the isomer shift for InSb and Cu$_2$Sb and thier 
various lithiated products is reported. The average insertion/intercalation voltage and the volume 
expansion for transitions from InSb to Li$_3$Sb and Cu$_2$Sb to Li$_3$Sb are calculated
and found to be in good agreement with the experimental values. These findings help to resolve the 
uncertainity regarding the lithiation mechanism in InSb.   
\end{abstract}

\maketitle

\section{INTRODUCTION}

Intermetallic compounds present an attractive alternative to graphite as anode materials in Li ion 
insertion batteries due in particular to the high capacity, an acceptable rate capability and operating 
potentials well above the potential of metallic lithium. 
The work in this direction gained tempo after the announcement that tin-based compounds 
\cite{idota97,mao99} could operate effectively as alloying hosts for Li ions.  In search of new anode 
materials one of the desirable qualities is that the structural and volumetric  
distortion of the host anode material on lithiation are small. This could be achieved if there existed 
a strong structural relation\cite{vaughey00,fransson01,kropf01} between the host and its lithiated products.
Some of the compounds which show such structural relations are Cu$_6$Sn$_5$
\cite{nordstrom01,benedek02,sharma03}, InSb\cite{vaughey00,johnson00,kropf01,hewitt01,vaughey01,tostmann02}, 
Cu$_2$Sb \cite{fransson01}, MnSb, Mn$_2$Sb\cite{fransson03}, SnSb \cite{li99}, GaSb\cite{vaughey01} etc. 
These compounds can be divided into three main structure types, namely 
NiAs (Cu$_6$Sn$_5$, MnSb), ZnS (InSb, GaSb) and Cu$_2$Sb (Mn$_2$Sb). 
In these materials, one of the components (Cu, In, Mn, Ga) is less active and is extruded on lithiation, 
while the other component provides the structural skeleton for further lithiation. 
In a previous work,\cite{sharma03} the lithiation of $\eta'$-Cu$_6$Sn$_5$ was studied. In the present work,
we investigate one representative compound from each of the other two structure types. These are InSb 
which belongs to the zinc blende structure and Cu$_2$Sb which is the prototype for the third class of 
materials.

Since it is desirable for an ideal anode material to reversibly insert/intercalate Li ions while
maintaining the structural stability, the knowledge of the exact amount of Li accommodated within the 
structure before the extrusion of the less active component is extremely important. In this regard, 
recently the electrochemistry of the Li insertion in InSb was studied experimentally, and contradictory 
results were published: 
1) Vaughey {\it et al.}\cite{vaughey00} initially proposed that 2 Li atoms could be incorporated 
within the InSb structure with very small volume expansion (5.6\%) before In extrusion, resulting in 
excellent capacity
2) Johnson {\it et al.}, \cite{johnson00} subsequently reported that In extrusion occurred at an earlier 
stage according to a general reaction: 
$(x+y){\rm Li + InSb } \rightarrow   {\rm Li}_{x+y}{\rm In}_{1-y}{\rm Sb}   
(0 \le x \le 2;  0 \le y \le 1)$ in which a 
Li$_y$In$_{1-y}$Sb zinc-blende framework played an important role in the Li insertion/In extrusion 
process, but they could not determine the precise amount of Li that could be inserted before the onset 
of In extrusion;
3) On the other hand, Hewitt {\it et al.} \cite{hewitt01} claimed, that $x$ reached a maximum value 
of 0.27 before In atoms were extruded to immediately form Li$_3$Sb in a two-phase reaction, and that 
Li$_2$Sb could be generated on subsequent electrochemical cycling;
4) From in situ X-ray absorption measurements, Kropf {\it et al.} \cite{kropf01} and 
Tostmann {\it et al.}\cite{tostmann02} provided supporting evidence that Li$_{x+y}$In$_{1-y}$Sb 
electrode compositions were generated during the initial discharge of Li/InSb cells and that 40\% of 
the extruded In was not incorporated back into the Sb lattice during the subsequent charging of the cells. 
In the light of the uncertainty of the reaction process, there is a need to clarify the exact lithiation 
path. To this extent we perform total-energy calculations for studying the mechanism of Li 
insertion/intercalation in InSb. A similar pathway is also provided for Cu$_2$Sb. We also present the 
theoretically determined isomer shift (IS) for these materials and various lithiated products. 

The change in the bonding character on lithiation is another important information that can
be extracted from the first-principles calculations. Such analysis regarding battery materials, 
has been done in the past using the total\cite{sharma03} or difference\cite{ven98,aydinol97} 
charge density. But it was noted by Bader {\it et al.}\cite{bader84} and Becke {\it et al.} 
\cite{becke90} that the information regarding the bonding in the molecular systems is not fully 
contained in the density ($\rho$) but rather in its Laplacian ($\nabla^2 \rho$) and in the kinetic 
energy density. These quantities are used to calculate the electron-localization-function (ELF), 
\cite{becke90} which varies between 0 and 1, with 1 corresponding to perfect localization and 1/2 
corresponding to electron gas like behaviour. Thus ELF in a way quantifies the bonding between various 
atoms in a solid and hence is ideal for studying the change in the bonding nature as new bonds are formed 
and (or) old bonds are weakened or broken by the insertion of Li atoms. We present the ELF for some of 
the lithiated products of InSb in the present work.

The paper is arranged in the following manner. In Section II, we present the details of the calculations. 
Section IIIA deals with the energetics of lithiating InSb and Cu$_2$Sb. To facilitate the study of the  
change in 
the bonding character on the lithiation of these compounds the IS and the ELF are given in Section IIIB.
Section IV provides a summary of our work.

\section{METHODOLOGY}

Total energy calculations are performed using the full potential linearized augmented planewave (FPLAPW) 
method. In this method, the unit cell is partitioned into non-overlapping muffin-tin spheres 
around the atomic sites and an interstitial region. In these two types of regions different basis sets 
are used. The scalar relativistic Kohn-Sham equations are solved in a self-consistent scheme.  
For the exchange-correlation potential we use the local density approximation (LDA). All the calculations 
are converged in terms of basis functions as well as in the size of the $k$-point mesh representing the 
Brillouin zone. 

In order to compare the total energies, calculated from first principles, for any two structures one has 
to optimize the structural parameters. These parameters are the atomic positions and the unit cell volume, 
which are varied until the global energy minimum is found. Thereby the experimental crystalline data is 
used as a starting point. For all the energy differences reported in the present work these optimized 
crystal structures are used. For these calculations the existing {\sf WIEN2k} code\cite{WIEN} is used. 

The details for the calculation of the IS of $^{121}$Sb have been presented before. \cite{sharma02} 
In the present work, the same value for the nuclear specific parameter ($\alpha = -0.846 \pm 0,002 a_0^3$ 
mm/s) is used to convert the contact charge densities to the IS. The effect of the muffin-tin sphere 
size on the IS has also been studied previously.\cite{sharma02} It was found that the 
IS is not affected by the choice of the muffin-tin radius. A value of 2.0a$_0$ has been used in the 
present work for all the atoms. 

The details for the ELF can be found in Ref. \onlinecite{becke90}. These calculations are performed 
using the newly developed {\sf EXCITING } code. Calculation of ELF in a mixed basis like LAPW requires 
the kinetic energy density to be made smooth across the muffin-tin sphere. Thus these calculations are 
performed using super-LAPW, \cite{singh91} which requires the matching of the wave function and its 
first and second derivatives at the muffin-tin sphere boundary making calculations computationally 
more demanding but affordable because the size of the unit cells is relatively small.  

\section {RESULTS AND DISCUSSION}

\subsection{Lithiation of InSb and Cu$_2$Sb}

InSb exists in the zinc blende structure with In atoms at (0,0,0) and the Sb atoms occupying four 
tetrahedral sites: (0.25,0.25,0.25),
(0.75,0.25,0.75), (0.75,0.75,0.25) and (0.25,0.75,0.75). This kind of structural arrangement leaves 
four tetrahedral sites vacant  namely (0.75,0.25,0.25), (0.25,0.25,0.75), (0.25,0.75,0.25) and 
(0.75,0.75,0.75). Along with this there is another void in the structure at (0.5,0.5,0.5).  
On lithiation of InSb (or other zinc blende compounds),  there are two  distinct probable processes: 
1) The incoming Li ions go to one of these voids. 
2) A substitutional reaction takes place, i.e., In (or the less active component) is extruded and 
is replaced by the incoming Li ions. In light of the existing controversy regarding these processes in 
InSb, we studied various possible lithiation paths. Using first-principles total-energy 
calculations we determine the formation energies for various lithiated products of InSb using
the following formula:
\begin{equation}
\rm{yLi + Li_xInSb \rightarrow Li_{x+y}In_{1-y}Sb + yIn}
\end{equation} 
The formation energy for the compound Li$_{\rm{x+y}}$In$_{\rm{1-y}}$Sb is given as
\begin{equation}
\Delta \rm{ E = E_{Li_{x+y}In_{1-y}Sb} + y E_{In}^{met} - [ E_{Li_xInSb}+ y E_{Li}^{met} ]}. 
\end{equation} 
Here E$_C$ is the total energy of the
compound $C$ and E$_A^{\rm{met}}$ is the energy per atom for metallic $A$. The formation energy defined 
this way reflects the relative stability of the compound  Li$_{\rm{x+y}}$In$_{\rm{1-y}}$Sb 
with respect to the reactants Li and Li$_{\rm y}$InSb.\cite{ven98} 

\begin{figure}[ht]
\centerline{\includegraphics[width=0.85\columnwidth,angle=-90]{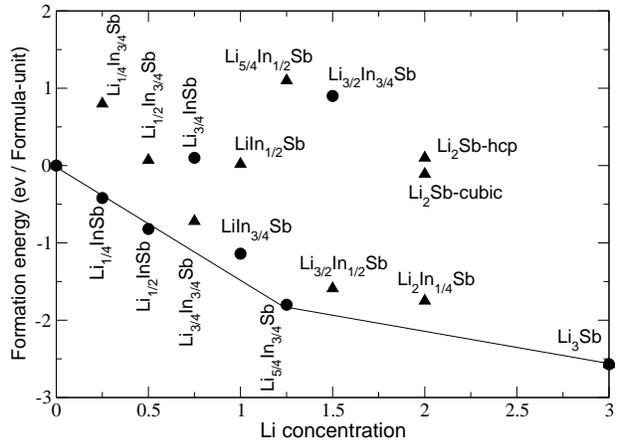}}
\caption {The formation energies in eV per formula unit. The compounds formed from the 
insertion/intercalation of Li atom in the previous compounds are represented by circles and compounds 
formed by the substitutional reaction are represented by triangles. }
\end{figure}

\begin{figure}[ht]
\centerline{\includegraphics[width=0.9\columnwidth,angle=-90]{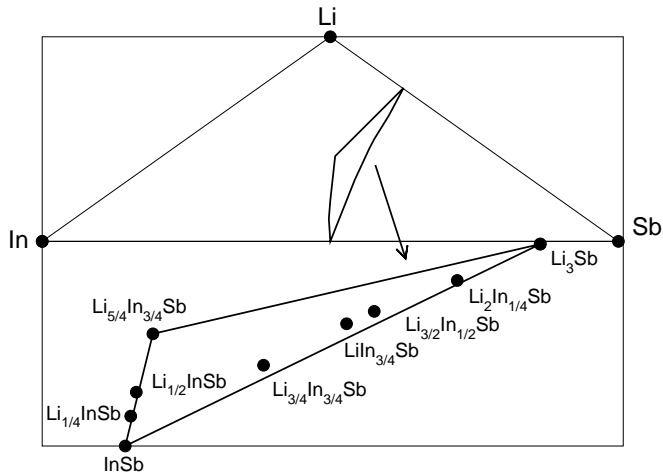}}
\caption {(a) A part of Li-In-Sb phase diagram. (b) Close up of the energetically favourable InSb-Li3Sb 
part of the phase diagram}
\end{figure}

These formation energies for InSb and its lithiated products are presented in Fig. 1. 
The compounds formed via the insertion/intercalation reaction of Li with the previous compound are shown 
as circles, while the compounds formed via the substitutional reaction are marked by triangles. 
As can be seen the formation energies of certain compounds are negative, {\it i.e.} these compounds 
are stable with respect to the reactants. A part of the Li-In-Sb phase diagram is shown in Fig. 2. 
Fig. 2(a) consists of the Gibb's trangle which consists of the energetically stable compounds 
in transition from InSb to Li$_3$Sb, which are 
shown in detail in Fig 2(b). We find that insertion of 1/2 Li 
atoms per InSb formula unit leads to a volume expansion of 5.3\%. In atoms are extruded beyond this 
concentration of Li, leading to very little further volume expansion. These results show best agreement 
with the experimental conclusions of Tostmann {\it et al.}\cite{tostmann02} The maximum number of 
intercalated Li atoms before In extrusion as reported by Hewitt {\it et al.}\cite{hewitt01} is almost 
half the value obtained in the present work.  
According to the energetics, the compound Li$_2$InSb is unstable, but if formed leads to an enormous 
volume expansion of 15.4\%. The total theoretical volume expansion on going from InSb to Li$_3$Sb is 
5.6\% which is slightly more than the experimentally reported value of 4.4\%. It can be noted from 
Fig. 1 that, like Hewitt{\it et al.},\cite{hewitt01} we find the formation of 
Li$_2$Sb during the first cycle to be energetically hindered. But we also note that the cubic phase of 
this compound is lower in energy than the hexagonal phase proposed earlier.\cite{hewitt01}

The initial incoming Li atoms have the freedom to occupy either 
the site (0.75,0.25,0.25) or (0.5,0.5,0.5) in the InSb framework to form Li$_{0.25}$InSb. 
We find that the probability for Li insertion at the position (0.75,0.25,0.25) is higher than at 
(0.5,0.5,0.5). The difference in energy between the two structures, Li$_{0.25}$InSb 
with Li at (0.75,0.25,0.25) and Li$_{0.25}$InSb with Li at (0.5,0.5,0.5), is 0.05 eV per formula unit.
 Our calculated average intercalation voltage (AIV) for a transition from InSb to Li$_3$Sb is 0.83V, 
which is higher than the experimental value of the plateau in the voltage profile 
(0.6V \cite{tostmann02} and 0.7V \cite{hewitt01}). 

Another important thing to be noted from Fig. 1 is the difference in the formation energies 
between the two consecutive compounds. After the formation of Li$_{5/4}$In$_{3/4}$Sb it is energetically 
less demanding for Li atoms to come in and (or) for In atoms to extrude from the structure. 
Also in the case of the reverse reaction it is energetically more demanding for In atoms to reincorporate 
in the structure beyond a concentration of 75\%. This is in accordance with the 
experimental findings of Tostmann {\it et al.},\cite{tostmann02} that only about  60\% of the In atoms 
go back into the structure on charging and that Li atoms are extracted more easily and rapidly from this 
In deficient phase of the kind $\rm{Li_{0.6+x}In_{0.4}Sb}$ leading to a good cycling capacity. 

\begin{figure}[ht]
\centerline{\includegraphics[width=0.85\columnwidth,angle=-90]{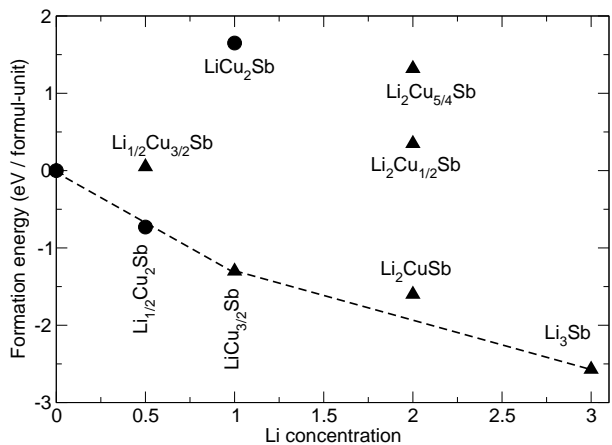}}
\caption {The formation energies in eV per formula unit. The compounds formed from the 
insertion/intercalation of Li atom in the previous compounds are represented by circles, and compounds 
formed by the substitutional reaction are represented by triangles. }
\end{figure}

\begin{figure}[ht]
\centerline{\includegraphics[width=0.9\columnwidth,angle=-90]{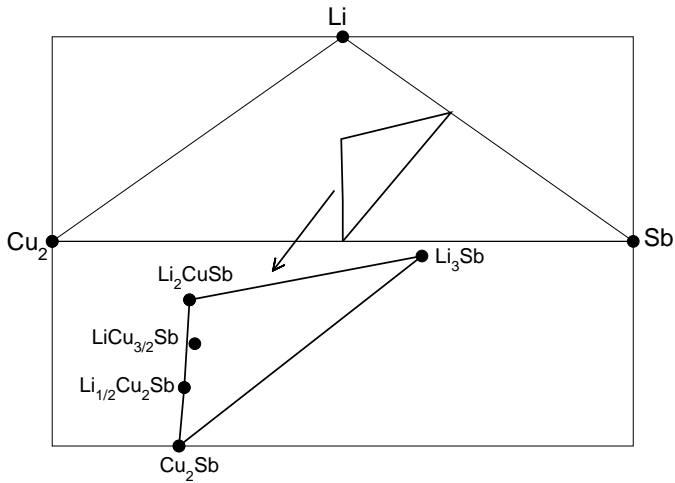}}
\caption {(a) A part of Li-Cu$_2$-Sb phase diagram. (b) Close up of the energetically favourable 
Cu$_2$Sb-Li3Sb part of the phase diagram}
\end{figure}

Cu$_2$Sb exists in a tetragonal structure with space group P4/nmm and 
2 formulae units per unit cell. The Cu atoms occupy the sites (0.75,0.25,0), (0.25,0.75,0.0), 
(0.25,0.25,0.27) and (0.75,0.75,0.73) while the Sb atoms are situated at (0.25,0.25,0.70) and 
(0.75,0.75,0.3). There exist voids in the structure at (0.5,0.5,0.5), (0.25,0.75,0.5) and (0.75,0.25,0.5). 
The two reaction mechanisms investigated in the present are,
1) the insertion/intercalation of Li atoms, and
2) the substitution of Cu by Li atoms. 
The formation energies for these reaction paths are presented in Fig. 3. Similar to the previous case, 
the Cu atoms are extruded from the structure after insertion of 1/2 Li atoms per Cu$_2$Sb formula unit. 
The first incoming Li atoms occupies the site (0.25,0.75,0.5) in the unit cell. Further lithiation leads 
to replacement of the Cu atom at site (0.75,0.25,0.0) by a Li atom to form LiCu$_{3/2}$Sb. 
The optimized structural parameters for LiCu$_{3/2}$Sb show a volume expansion of 18\% per formula unit 
compared to Cu$_2$Sb. More incoming Li atoms leads to the formation of the Li$_2$CuSb and a free Cu atom. 
Li$_2$CuSb exists in a cubic structure with space group F$\overline {4}$3m. In this unit cell the Sb atoms are present 
at the origin, the face centers and four of the tetrahedral sites i.e. (0.75,0.75,0.75), (0.25,0.25,0.75), 
(0.75,0.25,0.25) and (0.25,0.75,0.25) occupied by the Cu atoms. The Li atoms are located at the other 
four tetrahedral sites and at (0.5,0.5,0.5).  
The theoretically optimized structure of Li$_2$CuSb shows a volume expansion of 26.4\% per formula unit 
compared to the  initial compound Cu$_2$Sb. This is in good agreement with the experimental value of 
25.2\%.\cite{fransson01} 

 Energetically favourable Cu$_2$Sb to Li$_3$Sb part of the phase diagram is shown  in Fig. 4. 
The calculated value of the AIV for going from Cu$_2$Sb 
to Li$_3$Sb is 0.97 V, which is slightly higher than the experimental value of 0.82 V.\cite{fransson01}

\subsection{Isomer shift and electron localization function}

\begin{figure}[ht]
\centerline{\includegraphics[width=0.85\columnwidth,angle=-90]{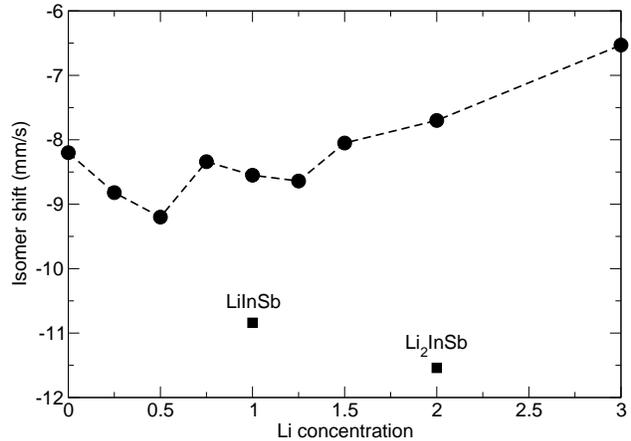}}
\caption {Isomer shift in mm/s for the compounds on the energetically favourable lithiation reaction paths 
for InSb.}
\end{figure}

An excellent method of studying the change in the local environment and symmetry is the M\"ossbauer IS. 
We have calculated the $^{121}$Sb IS of various lithiated products of InSb and the results are presented 
in Fig. 5. The calculated value for InSb is in good agreement with the experimental data of -8.6 mm/s. 
\cite{pruitt70} Insertion reaction of the Li atoms with the host makes the IS steadily more negative with 
a sudden increase when the In is extruded from the structure. The IS for the energetically hindered 
compounds LiInSb and Li$_2$InSb are -10.54 mm/s and -11.54 mm/s respectively. These values are largely 
negative as compared to the other lithiated products of InSb making them easily identifiable by M\"ossbauer 
spectroscopy. 

\begin{figure}[ht]
\centerline{\includegraphics[width=0.85\columnwidth,angle=-90]{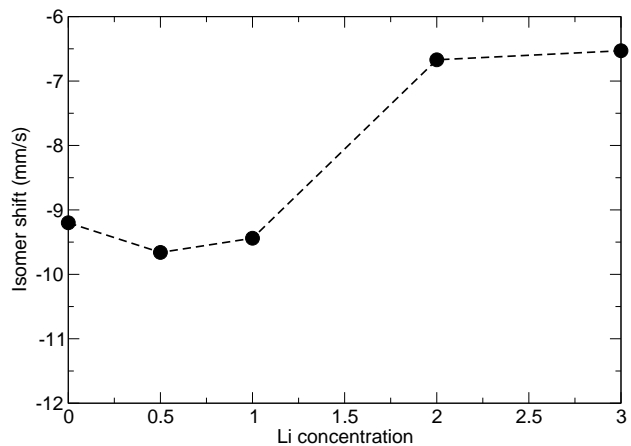}}
\caption {Isomer shift in mm/s for the compounds on the energetically favourable lithiation reaction paths 
for Cu$_2$Sb.}
\end{figure}

It can be seen from Fig. 6, the IS for Cu$_2$Sb, like InSb, shows a decrease on lithiation via the 
insertion reaction and the continuous increase when the Cu extrusion starts. The IS has been shown to 
depend linearly on the partial number of $s$ electrons (N$_s$)\cite{sharma02} and the variation of the 
IS with the Li concentration can be explained very well using this fact. As more Li atoms 
insert/intercalate in the host material, a part of the charge lost by them is gained by the Sb atoms. 
This causes an increase in N$_s$ and hence a more negative IS. On the other hand, on lithiation via 
the substitution reaction there is no excess charge. 
This causes the decrease in N$_s$ and hence the IS becomes less negative. Thus the change in the IS with 
the Li atom concentration is a good signature of the reaction mechanism and hence we suggest such 
experiments to be performed to shed more light on the reaction path and for the unambiguous clarification 
of whether compounds like LiInSb and Li$_2$InSb are formed or not. 

\begin{figure}[ht]
\centerline{\includegraphics[width=0.85\columnwidth,angle=-90]{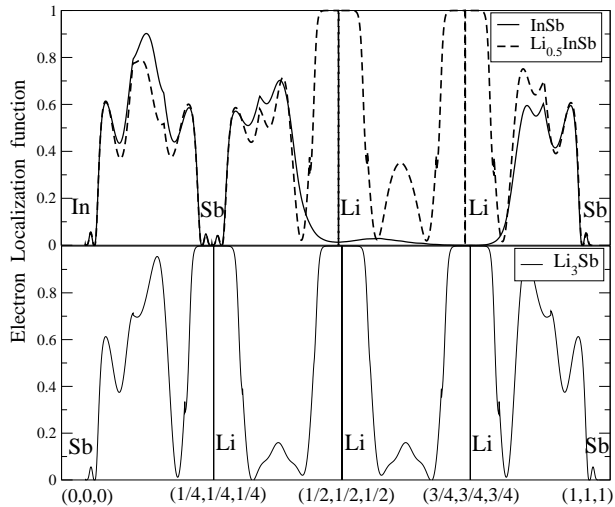}}
\caption {Electron localization function for InSb, Li$_{0.5}$Sb and Li$_3$Sb.}
\end{figure}

The change in the bonding character on lithiation of the anode material can be studied using the
charge density plots which have been used for Cu$_6$Sn$_5$ \cite{sharma03} and Cu$_2$Sb 
\cite{sharmaPP}. In the present work we use ELF for such an investigation of InSb. Fig. 8 shows the
ELF for InSb, Li$_{0.5}$InSb and Li$_3$Sb along the (111) direction. A covalent character of bonding 
between In atom at (0,0,0) and Sb atom at (1/4,1/4,1/4) can be seen as a strong localization of
charge between the two atoms. Since only the valence electrons participate in the bond formation, 
the core
charge has been ignored for all the atoms. The high value of ELF at the Li atom site is due the core
states being taken into consideration as local orbitals. Lithiation of InSb to form  Li$_{0.5}$InSb
leads to weakening of the bonding between In and Sb. This manifests itself by a lowering of the
ELF from 0.9 to 0.79 between the two atoms. The bond between Li and Sb atoms is ionic in nature.
This ionocity is retained all the way upto the formation of  Li$_3$Sb. The presence of delocalized
electrons between atoms is indicative of the metallic nature of the compound.

\section{Summary}


To summarize, the two prototype structures, InSb and Cu$_2$Sb, for the use as anode materials in 
Li-ion batteries are investigated in the present work. From first-principles total-energy calculations 
we conclude the following:
\begin{itemize}
\item On lithiation of InSb during discharge, the In atoms are extruded from the structure after 
insertion/intercalation of half a Li atom per InSb formula unit. The formation of Li$_{1/2}$InSb leads to 
a volume expansion of 5.4\%
\item It is much easier for In and Li atoms to come out of and go into the structure after formation of 
Li$_{5/4}$In$_{3/4}$Sb which supports the good cycling capacity. 
\item The formation of the compounds  LiInSb, Li$_2$InSb and Li$_2$Sb is energetically hindered. 
In particular, the formation of LiInSb or Li$_2$InSb would lead to enormous volume expansions of 14.4\% 
and 15.4\%, respectively, which is much more than any experimental observation. 
\item Our calculated AIV for the transition from InSb to Li$_3$Sb is in agreement with the 
experimental value of the plateau in the measured voltage profile. These findings make our results 
closest to that of Tostmann {\it et al.}\cite{tostmann02}
\item In the case of Cu$_2$Sb, the Cu atoms are extruded after insertion of half a Li atom per Cu$_2$Sb 
formula unit. 
\item The trends in the IS with increasing Li concentration is an excellent signature of the 
lithiation reaction mechanism. The formation of compounds like LiInSb and Li$_2$InSb can be unambiguously 
determined using M\"ossbauer spectroscopy, since these compounds have large negative IS values compared 
to any other lithiated product of InSb. 
\item The ELF shows that lithiation of InSb results in the weakening of bonds between 
In and Sb atoms which would lead, upon further lithiation, to an extrusion of In atoms. The bonding 
changes from covalent for InSb to ionic in nature for Li$_3$Sb.  

\end{itemize}

\noindent \textbf{Acknowledgements}

We thank the Austrian Science Fund (project P16227) and the EXCITING network funded by the EU (Contract HPRN-CT-2002-00317)
for  financial support.
SS would also like to thank K. Edstr\"om and  M. M. Thackeray for useful discussions and suggestions during the course of this work.


\end{document}